# Direct and indirect magnetocaloric effects near room temperature related to structural transitions in $Y_{0.9}Pr_{0.1}Fe_2D_{3.5}$ deuteride

V. Paul-Boncour[a,*], A. Herrero[a,1], V. Shtender[a,2], K. Provost[a], E. Elkaim[b]

[a] Université Paris Est Créteil, CNRS, ICMPE, 2 Rue Henri Dunant F-94320 Thiais, France

[b] SOLEIL, L'Orme des Merisiers, Saint-Aubin, BP 48, 91192 Gif-sur-Yvette Cedex, France

**Abstract**

The structural and magnetic properties of $Y_{0.9}Pr_{0.1}Fe_2D_{3.5}$ deuteride have been investigated by synchrotron and neutron diffraction, magnetic measurements, and differential scanning calorimetry. Deuterium insertion induces a 23.5 % cell volume increase and a lowering of crystal symmetry compared to the cubic C15 parent compound (*F*d-3*m* SG). The deuteride is monoclinic (*P*2$_1$/c SG) below 330 K and undergoes a first order transition between 330 and 350 K towards a pseudo-cubic structure (*R*-3*m* SG) with $T_{O\text{-}D}$ = 342(2) K. The compound is ferromagnetic, accompanied by a magnetostrictive effect below $T_C$ = 274 K. The analysis of the critical exponents indicates a second order type transition with a deviation from the isotropic 3D Heisenberg model towards the 3D XY model. This implies an easy plane of magnetization in agreement with cell parameter variation showing a planar magnetic orientation. A weak magnetic peak is even observed at the order-disorder transition with a maximum at 343 K. Magnetic entropy variations are characteristic of direct and reverse magnetocaloric effects at $T_C$ and $T_{O\text{-}D}$ respectively.

* Corresponding authors: tel :+33 1 49 78 12 07, email: paulbon@icmpe.cnrs.fr

Present address:

[1]Departamento de Física Aplicada I, Escuela de Ingeniería de Bilbao, Universidad del País Vasco UPV/EHU, Plaza Torres Quevedo 1, 48013, Bilbao, Spain

[2]Department of Chemistry – Ångström Laboratory, Uppsala University, Box 538, 75121 Uppsala, Sweden

# 1. Introduction

$R$Fe$_2$ Laves phase compounds ($R$ = rare earth) have been investigated for their remarkably large magnetic anisotropy and magnetostriction up to room temperature [1, 2]. For example, the Tb$_{0.3}$Dy$_{0.7}$Fe$_{1.93}$ compound (Terfenol-D) has been developed and its composition optimized for applications in magneto-mechanical sensors and actuators, as acoustic and ultrasonic transducers [3-6].

These $R$Fe$_2$ compounds have also raised interest for the modification of their structural and magnetic properties upon hydrogen insertion [7-14]. They can absorb up to 5 H/f.u. and their pressure composition isotherms (PCI) show a multi-plateau behavior, which has been explained by the existence of hydrides with different structures [15-19]. All these structures are due to various distortions and/or superstructures of the cubic C15 type parent compound depending on the H content [20-27]. The hydrogen (deuterium) atoms are mainly located into tetrahedral $R_2$Fe$_2$ and $R$Fe$_3$ interstitial sites as observed by neutron diffraction. This yields a displacement of the metal atoms and a long-range ordering of the H(D) atoms in the structure at room temperature. Structural order-disorder transitions from ordered towards disordered cubic structure are generally observed upon heating and occur at a $T_{O-D}$ temperature which depends on the H content [24, 28, 29]. Hydrogen insertion in these $R$Fe$_2$ compounds yields a decrease of the Curie temperature $T_C$ and large changes of the Fe-Fe and $R$-Fe interactions because of both structural variations and modification of the density of state (DOS) near the Fermi level ($E_F$) [30].

Many studies have been already performed on YFe$_2$-H$_2$ system and a structural and magnetic phase diagram proposed [29]. YFe$_2$D$_x$ compounds with $x$=1.2, 1.75 and 1.9 display tetragonal and cubic superstructures below $T_{O-D}$ [24, 31]. Crystal structure of YFe$_2$D$_{3.5(1)}$ deuteride at room temperature was first described in a rhombohedral lattice [22, 32], but further neutron powder diffraction (NPD) studies showed a monoclinic distortion associated with a superstructure [33]. A transition towards a rhombohedral structure was observed at $T_{O-D}$ = 350 K. YFe$_2$D$_{3.5(1)}$ has a ferromagnetic behavior with $T_C$= 310-360 K depending on the references [25, 32, 34-36], probably due to slightly different H(D) contents or analysis method.

YFe$_2$(H,D)$_{4.2(1)}$ compounds crystallize in a monoclinic structure below $T_{O-D}$. A first order ferromagnetic-antiferromagnetic transition (FM-AFM) was observed at $T_{FM-AFM}$ = 84 K for the deuteride and 131 K for the hydride [28, 29, 37]. Beside this giant isotopic behavior, a direct magnetocaloric effect [38] was observed at this transition accompanied by a cell volume change [39]. YFe$_2$D$_5$ is orthorhombic and no clear O-D transition has been observed [40]. It behaves as a Pauli paramagnet down to low temperature. Band structure calculations have confirmed the paramagnetic ground state of the Fe sublattice due to strong Fe-H bonding [41, 42].

Several studies have been undertaken to increase $T_{\text{FM-AFM}}$ and get an MCE near room temperature (RT) for magnetic refrigeration applications by playing on Y substitution in YFe$_2$H$_{4.2}$ hydride. For $R$= Gd, Tb, which have a larger atomic radius than Y, the full replacement allows to drive the maximum magnetic ordering temperature near only 200 K, with a reduced MCE [39, 43-46].

Another solution to observe an MCE near RT is to adjust the H content. As previously indicated YFe$_2$H$_{3.5}$ displays a Curie temperature between 308 and 350 K. A partial substitution of Y by a rare earth with larger radius should allow to tune $T_C$. Beside a larger radius than Y, light $R$ elements have also the particularities to have a ferromagnetic coupling with Fe, contrary to heavy $R$ display a ferrimagnetic ground state. We expect therefore an improved MCE effect near RT by substituting Y by Pr.

For these reasons, we have studied in details Y$_{0.9}$Pr$_{0.1}$Fe$_2$D$_{3.5}$ deuteride by determining its crystal and magnetic structures. Its chemical composition was chosen as it is well adapted for neutron diffraction study (Pr and D have moderate neutron absorption and weak inelastic contribution compared to Sm, Gd and H). We have combined neutron powder diffraction (NPD) experiments on two different spectrometers with synchrotron radiation X-ray diffraction (SR-XRD) measurements to solve its structure at RT and follow its evolution versus temperature. The phase transitions were also investigated by differential scanning calorimetry (DSC) and the magnetic and magnetocaloric properties by magnetic measurements. Our main purpose was to investigate magnetostrictive effects and order-disorder transitions to determine whether they are correlated and observe the corresponding magnetocaloric effects near room temperature. These results will be compared and discussed with previous works on $R$Fe$_2$ hydrides.

## 2. Experimental methods

Y$_{0.9}$Pr$_{0.1}$Fe$_2$ intermetallic compound was prepared by induction melting of the pure elements (Y 99.9 %, Pr 99.9 %, Fe 99.99 %) followed by three weeks annealing treatment under vacuum at 1100 K. The chemical composition of the alloy measured by electron probe microanalysis (EPMA) was Y$_{0.91(1)}$Pr$_{0.09(1)}$Fe$_{1.98(2)}$. The analysis of the X-ray powder diffraction (XRD) pattern of the alloy indicates that it contains 98.7 wt % of the C15 cubic phase ($a$ = 7.36702(5) Å), 0.9 wt % of $Y_2O_3$ and 0.4 wt % of $Y_6Fe_{16}O$. $Y_{0.9}Pr_{0.1}Fe_2D_{3.5}$ deuteride was prepared by solid-gas reaction using a Sieverts apparatus. It was activated one night under primary vacuum at 400 K and hydrogenated at RT with the final pressure of 0.2 bar $D_2$. The deuteride was quenched into liquid nitrogen followed by a slow heating under air to poison the surface and prevent fast gas desorption.

The alloy and the deuteride were characterized by XRD at room temperature with a Bruker D8 diffractometer operating with a Cu-K$\alpha$ wavelength. Synchrotron radiation measurements (SR-XRD) were performed on the 2-circle diffractometer using the CRISTAL beam line at SOLEIL (Saint Aubin, France). The detection system was made of 9 linear Mythen modules covering a 50° $2\theta$ circle arc at 720 mm from the sample. The wavelength refined with a $LaB_6$ reference sample was $\lambda = 0.51302$ Å and the $2\theta$ range expanded from 1.2 to 52 °. The powder sample was placed in a sealed glass capillary tube (0.3 mm diameter), which was rotated to ensure homogeneity. The sample was cooled or heated using a gas streamer cooler operating from 130 K up to 360 K. NPD patterns of $Y_{0.9}Pr_{0.1}Fe_2D_{3.5}$ have been recorded at the Laboratoire Léon Brillouin (LLB, CEA, Saclay, France) at 2, 300 and 345 K on the high resolution 3T2 spectrometer with a wavelength of 1.225 Å, between 15° and 125° with a step of 0.05° and at selected temperatures from 1.5 to 345 K on the G4.1 spectrometer with a wavelength of 2.422 Å between 10 and 90° with a step of 0.1°. All the diffraction patterns were refined with the Fullprof code [47]. The lowering of crystal symmetry was investigated using Subgroup-graph from Bilbao server and Powdercell (PCW) codes. Location of D atoms was studied with the Free Object for Crystallography (FOX) code.

The thermal transitions were measured using a DSC Q100 from TA instrument, with a heating /cooling rate of 10 K/min between 100 and 600 K. Magnetic measurements were performed from 10 to 300 K using a conventional Physical Properties Measurement System (PPMS) from Quantum Design with maximum field of 9 T. Further magnetic curves were measured from 300 to 360 K with a DSM8 Manics magnetometer operating with a maximum field of 1.7 T.

The thermal dependence of the isothermal magnetic entropy change $\Delta S_M$ was calculated from magnetization isotherms recorded on field increase by numerical integration of one of the thermodynamic Maxwell's relations $\Delta S_M(T) = \mu_0 \int_0^H \left( \frac{\partial M(H,T)}{\partial T} \right)_H dH$ as described in [48].

The $-\Delta S_M(T)$ was evaluated in two regions of interest: around the FM transition located at 274 K and around a weak peak in magnetization found at T = 343 K. For the former region isotherms around $T_C$ were measured with temperature increments of 3 K and field changes $\mu_0\Delta H$ up to 5 T, with field steps of 0.2 T. For the latter case, isotherms were collected with steps of 2 K in the vicinity of T = 343 K and 5 K in further regions. The maximum field change in this case was $\mu_0\Delta H$=1.5 T.

## 3. Results

*3.1 X-ray and neutron diffraction*

*3.1.1 Crystal structure at 300 K*

A comparison of the powder diffraction patterns for $Y_{0.9}Pr_{0.1}Fe_2D_{3.5}$ measured at 300 K on CRISTAL synchrotron beam line, 3T2 and G4.1 neutron spectrometers is illustrated in Fig. 1, showing the complementarity of the different measurements. At low $Q$ different groups of diffraction lines are observed by SR-XRD ($Q$ = 2.2-2.4 Å$^{-1}$) and NPD ($Q$= 1.2-1.7 Å$^{-1}$) due to the difference of interaction with matter. The influence of the instrumental resolution is clearly seen between neutron and synchrotron diffraction, but also between the two neutron spectrometers. G4.1 has a better resolution at low angle than 3T2 but expends to a more limited $Q$ range (0.5-3.7 Å$^{-1}$ for G4.1 and 0.5-9 Å$^{-1}$ for 3T2).

The SR-XRD pattern measured at 300 K was refined with a main phase indexed in a monoclinic structure, few percent of secondary phases due to oxides ($Y_2O_3$ and $Y_6Fe_{16}O$) which were already present in the initial alloy. Assuming that the monoclinic structure is due to a distortion of the C15 cubic structure, the monoclinic space group (SG) is obtained by a successive lowering of crystal symmetry from *Fd-3m* (227) to *R-3m* (166) and then *C*2/*m* (N°12) SG. The monoclinic parameters measured in both laboratories and synchrotron experiments remain very close to each other (Table 1). The refined synchrotron pattern is shown in Fig. 2 (a). Compared to the parent alloy, a cell volume augmentation of 23.5 % is observed corresponding to a swelling of 3.35 Å/ D atom.

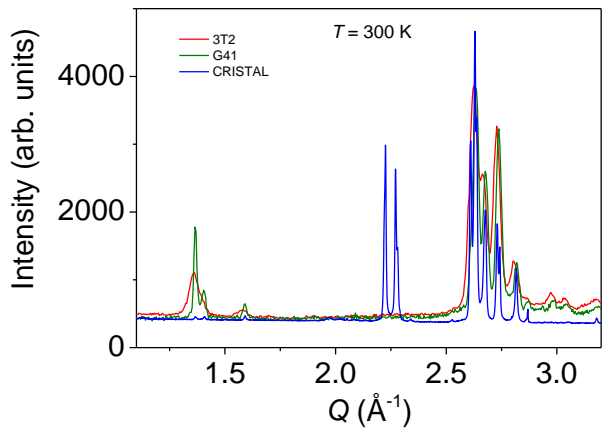

**Fig. 1**: Comparison of the three patterns for $Y_{0.9}Pr_{0.1}Fe_2D_{3.5}$ measured at 300 K on CRISTAL, 3T2 and G4.1.

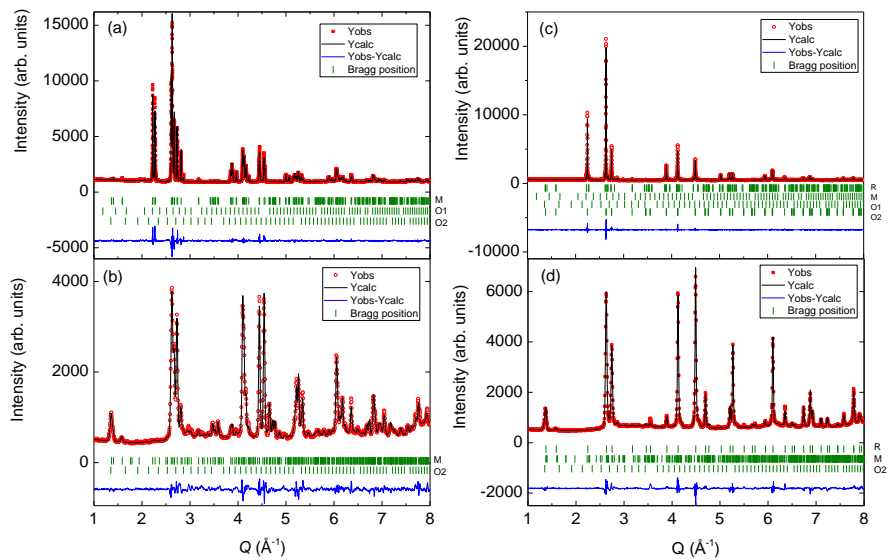

**Fig. 2 :** Refined SR-XRD patterns of $Y_{0.9}Pr_{0.1}Fe_2D_{3.5}$ measured on CRISTAL ($\lambda = 0.51302$ Å) at 300 K (a) and 346 K (c); refined NPD patterns measured on 3T2 ($\lambda = 1.225$ Å) at 300 K (b) and 345 K (d). (R) Rhombohedral, (M) Monoclinic, (O1) $Y_2O_3$ and (O2) $Y_6Fe_{16}O$ phases.

The 3T2 NPD pattern was first refined in $C2/m$ space group. The lowering of crystal symmetry generates 7 $R_2Fe_2$ sites and 3 $RFe_3$ sites ($R=$ Y, Pr). A good refinement is obtained with 7 over 10 sites partially occupied by D atoms (5 $R_2Fe_2$ sites and 2 $RFe_3$ sites) and a total number of D atoms $n_D= 4.0(2)$ D/f.u., which is slightly larger than the nominal content. Note, that $Y_2O_3$ is not observed in NPD compared to the SR-XRD pattern. This can be due to the relatively lower surface/bulk contribution indicating that the oxide species should be rather located at the surface of the grain and related to the surface poisoning treatment.

However, few weak diffraction peaks in the range 3 - 3.6 Å$^{-1}$ were not indexed by $C2/m$ SG indicating that the ordering of D atoms into interstitial sites induces a further lowering of the crystal symmetry. These peaks can be indexed in a primitive monoclinic space group. According to group-subgroup scheme tree, the four different primitive subgroups of $C2/m$ are: $P2/m$ (10), $P2_1/m$ (11), $P2/c$ (13) and $P2_1/c$ (14). Refinement in a pattern matching mode, shows that one peak at $Q = 1.4$ Å$^{-1}$ is not indexed with $P2/c$ SG, whereas $P2_1/m$ and $P2/c$ SG generate peaks which are absent from the pattern below 1.4 Å$^{-1}$. Therefore, the most appropriate primitive SG to refine the NPD pattern is $P2_1/c$ (14). To confirm this assumption several tests have been performed to refine both G4.1 and 3T2 NPD patterns using the FOX code with all possible SG and only description in $P2_1/c$ SG converges to a solution. Additional trials considering a doubling of the $a$, $b$ or $c$ axis did not improve the NPD pattern refinement.

**Table 1**: Cell parameters of Y$_{0.9}$Pr$_{0.1}$Fe$_2$D$_{3.5}$ measured at 300 K on the different instruments.

| Experiment | D8-bruker | CRISTAL | 3T2 | 3T2 | G41 |
|---|---|---|---|---|---|
| λ (Å) | 1.54178 | 0.51302 | 1.225 | 1.225 | 2.422 |
| S.G. | $C2/M$ | $C2/M$ | $C2/M$ | $P2_1/c$ (setting 1) | $P2_1/c$ (setting 1) |
| $a$ (Å) | 9.5169(6) | 9.51568(7) | 9.5081(3) | 7.9069(3) | 7.9010(8) |
| $b$ (Å) | 5.65895(4) | 5.65968(4) | 5.6500(2) | 5.6506(2) | 5.6105(7) |
| $c$ (Å) | 5.5128(4) | 5.51279(4) | 5.5292(3) | 9.50792(2) | 9.4938(8) |
| $\beta$ (°) | 123.760 (3) | 123.7605(3) | 123.770(3) | 144.461(2) | 144.518(3) |
| $V$ (Å$^3$) | 246.83(3) | 246.829(3) | 246.92(2) | 246.92(1) | 244.28(4) |
| Deuteride (wt %) | 95.6(7) | 98.5(3) | 99.5(1) | 99.1(1) | 98.9(9) |
| Y$_2$O$_3$ wt % | 1.3( 1) | 0.3( 1) | - | - | - |
| Y$_6$Fe$_{16}$O wt % | 3.1( 2) | 1.2( 1) | 0.5(1) | 0.9(1) | 1.1(2) |
| R$_B$ (%) phase1 | 8.44 | 4.79 | 4.14 | 4.70 | 2.57 |
| $\chi^2$ | 1.67 | 9.51 | 7.68 | 7.33 | 2.74 |

There are 14 different settings for this SG and setting 1, was chosen as suitable for Fullprof code refinement, but gives a different description of cell parameters compared to $C2/m$. The refined D atomic positions and occupation numbers are reported in Table 2 and the refined 3T2 pattern is displayed in Fig. 2(b). The lowering of crystal symmetry in $P2_1/c$ SG generates 12 $R_2$Fe$_2$ (4$e$) sites and 4 $R$Fe$_3$ (4$e$) sites. The occupation numbers and then the atomic positions were refined and few D sites were found empty and removed from the refinement. Finally, 10 $R_2$Fe$_2$ and 1 $R$Fe$_3$ sites were found partially occupied, with only 4% occupation by D atoms in this last site. The total number of D atoms is 3.9(3) D/f.u., is still larger than obtained from the volumetric method (3.5(1) D/f.u.) but remains close within the standard deviation. Comparison of the different structures is displayed in Fig. S1.

**Table 2**: Atomic positions, occupancy factors and Debye Waller factor in $P2_1/c$ S.G.

| Atom | Wyckoff site | D site | x | y | z | $B$ (Å$^2$) | $N_{Occ}$ (/f.u.) | Occ % |
|---|---|---|---|---|---|---|---|---|
| Y/Pr | 4e | | 0.129(2) | -0.011(2) | 0.258(1) | 0.32(7) | 0.9/0.1 | 90/10 |
| Fe1 | 2b | | 0.5 | 0 | 0 | 0.77(5) | 1 | 100 |
| Fe2 | 2c | | 0 | 0 | 0.5 | 0.77(5) | 1 | 100 |
| Fe3 | 4e | | 0.500(2) | 0.243(2) | 0.756(2) | 0.77(5) | 1 | 100 |
| D1 | 4e | $R_2$Fe$_2$ | 0.478(2) | 0 | 0.336(2) | 0.6(1) | 0.91(3) | 91 |
| D2 | 4e | $R_2$Fe$_2$ | 0.921(8) | 0.738(7) | 0.035(8) | 0.6(1) | 0.26(3) | 26 |
| D3 | 4e | $R_2$Fe$_2$ | 0.101(7) | 0.741(6) | 0.960(5) | 0.6(1) | 0.30(3) | 30 |
| D4 | 4e | $R_2$Fe$_2$ | 0.839(7) | 0.762(5) | 0.927(5) | 0.6(1) | 0.34(3) | 34 |
| D5 | 4e | $R_2$Fe$_2$ | 0.130(5) | 0.830(5) | 0.028(5) | 0.6(1) | 0.33(2) | 33 |
| D6 | 4e | $R_2$Fe$_2$ | 0.145(4) | 0.058(11) | 0.751(8) | 0.6(1) | 0.15(2) | 15 |
| D7 | 4e | $R_2$Fe$_2$ | 0.890(4) | 0.120(4) | 0.265(4) | 0.6(1) | 0.43(3) | 43 |
| D8 | 4e | $R_2$Fe$_2$ | 0.686(10) | 0.660(7) | 0.518(7) | 0.6(1) | 0.22(2) | 22 |
| D9 | 4e | $R_2$Fe$_2$ | 0.515(6) | 0.226(4) | 0.449(5) | 0.6(1) | 0.44(3) | 44 |
| D10 | 4e | $R_2$Fe$_2$ | 0.460(5) | 0.240(4) | 0.542(4) | 0.6(1) | 0.40(3) | 40 |
| D11 | 4e | $R$Fe$_3$ | 0.425(9) | 0 | 0.667(7) | 0.6(1) | 0.14(2) | 14 |
| $N_{occ}$ total | | | | | | | 3.9(3) | |

All the interatomic distances were analyzed. Each Fe atom is surrounded by 6 Fe atoms at distances between 2.76 and 2.83 Å and D atoms at distances varying between 1.45 and 2.08 Å. Some calculated D-D distances are significantly shorter than the repulsive distance of 2.1 Å, between two D atoms [49, 50], but as all interstitial sites are partially occupied, this means that locally the D-D distances can remain larger than 2.1 Å. The existence of short range order between D atoms has been clearly observed in Laves phase deuterides such as ZrCr$_2$D$_4$ and YFe$_2$D$_{4.2}$ and analyzed with a Pair Distribution Function (PDF) [51, 52]. The comparison of the SR-XRD and NPD patterns in Fig. 1, clearly shows a background increase for the NPD patterns above 2.5 Å$^{-1}$.

*3.1.2 Crystal structure at 345 K*

A high temperature disordered phase (HT phase) has been observed in $Y_{0.9}Pr_{0.1}Fe_2D_{3.5}$. The 3T2 NPD pattern was measured at 345 K (Fig. 2(c)). The corresponding SR-XRD pattern, measured at 346 K was refined with a mixture of cubic ($Fd$-$3m$ SG) and monoclinic structure ($C2/m$ SG). The results are reported in Table 3, indicating 85 % of cubic phase, 13 % of monoclinic phase and 2 wt% of $Y_6Fe_{16}O$. The NPD pattern was first refined with the same cubic structure including additional D atoms. This yields 82 % of cubic phase, 16.7 % of monoclinic phase and 1.3 % of $Y_6Fe_{16}O$. The D atoms were mainly located in $R_2Fe_2$ sites (3.4 D f.u.) and few in the $R$Fe$_3$ sites (0.13 D/f.u.). However, the intensity of several cubic peaks was not very well fitted, in particular those with (hhh) Miller indices, and few peaks ($Q$ =1.59, 3.55, 7.93 Å$^{-1}$) were not indexed. This suggests a lowering of crystal symmetry and we have tested two different subgroups ($R$-$3m$ and $F23$ SG) according to previous publications on $YFe_2D_{3.5}$ and $ErFe_2D_{3.1}$ [22, 53]. All additional peaks can be indexed with both space groups.

**Table 3**: Structural parameters obtained from the refinement of SR-XRD and NPD patterns of $Y_{0.9}Pr_{0.1}Fe_2D_{3.5}$ at 345(1) K.

| Experiment | CRISTAL (346 K) | 3T2 (345 K) | 3T2 (345 K) |
|---|---|---|---|
| Phase 1 | | | |
| SG | $Fd$-$3m$ | $Fd$-$3m$ | $R$-$3m$ |
| $a$ (Å) | 7.90928 (3) | 7.9078(1) | 5.5923(1) |
| $c$ (Å) | | | 13.6966(6) |
| $V$ (Å$^3$) | 494.779( 3) | 494.51(1) | 370.96(2) |
| $V/Z$ (Å$^3$/f.u.) | 61.847(1) | 61.813(1) | 61.827(3) |
| wt % | 85.6(3) | 82.0 (1) | 82.7 (1) |
| $R_B$ | 4.60 | 5.61 | 3.88 |
| Phase 2 | | | |
| SG | $C2/m$ | $P2_1/c$ | $P2_1/c$ |
| $a$ (Å) | 9.5250(2) | 7.773(5) | 7.785(5) |
| $b$ (Å) | 5.6480(2) | 5.504(4) | 5.498(4) |
| $c$ (Å) | 5.5179(1) | 9.624(9) | 9.581(8) |
| $\beta$ (°) | 123.877(2) | 144.36(3) | 144.03(4) |
| $V$ (Å$^3$) | 246.45(1) | 239.9 (3) | 240.9(3) |
| $V/Z$ (Å$^3$) | 61.61(1) | 59.98(1) | 60.23(1) |
| wt % | 12.5( 2) | 16.7 (1) | 15.5 (1) |
| $R_B$ (%) | 14.8 | 17.8 | 6.83 |
| $Y_6Fe_6O$ wt % | 1.64(5) | 1.4 (1) | 1.8 (1) |
| $Y_2O_3$ wt % | 0.28(5) | - | - |
| $\chi^2$ | 11.4 | 8.1 | 5.2 |

However, the best refinement ($R_B$ =3.9 %) is obtained with the $R$-$3m$ space group as proposed for YFe$_2$D$_{3.5}$ [22] with 95.5 % D atoms located in 3 $R_2$Fe$_2$ sites and 4.5 % in 1 $R$Fe$_3$ site. The atomic parameters are given in Tables 3 and 4 and the refined pattern in presented in Fig. 2(d). The fit of the monoclinic phase is also improved with this rhombohedral SG as observed on $R_B$ values. As the rhombohedral cell parameters shows no elongation or contraction along the (111) axis within the instrumental synchrotron resolution, the cell is not distorted and the lowering of crystal symmetry is due to the partial ordering of the D atoms, leading to a loss of some symmetry elements as the four fold axis. It can be therefore considered as a pseudo-cubic cell.

**Table 4**: Atomic positions of the main rhombohedral phase ($R$-$3m$ SG) in Y$_{0.9}$Pr$_{0.1}$Fe$_2$D$_{3.5}$ at 345 K

| Atom | Wyckoff site | D site | $x$ | $y$ | $z$ | $B$(Å$^2$) | $N_{Occ}$ (atom/f.u.) | % Occ |
|---|---|---|---|---|---|---|---|---|
| Y/Pr | 6$c$ | | 0 | 0 | 0.1241(4) | 1.08(3) | 0.9/0.1 | 90/10 |
| Fe1 | 3$b$ | | 0 | 0 | 0.5 | 0.97(2) | 0.5 | 100 |
| Fe2 | 9$e$ | | 0.5 | 0 | 0 | 0.97(2) | 1.5 | 100 |
| D1 | 36$i$ | $R_2$Fe$_2$ | 0.228(1) | 0.304(1) | 0.4479(5) | 0.28(7) | 1.88(4) | 16 |
| D2 | 18$h$ | $R_2$Fe$_2$ | 0.8670(8) | 0.1330(8) | 0.2525(6) | 0.28(7) | 0.89(3) | 30 |
| D3 | 18$h$ | $R_2$Fe$_2$ | 0.828(1) | 0.172(1) | 0.290(1) | 0.28(7) | 0.47(2) | 16 |
| D4 | 6$c$ | $R$Fe$_3$ | 0 | 0 | 0.701(2) | 0.28(7) | 0.15(1) | 15 |
| $N$ (D/f.u.) | | | | | | | 3.4(1) | |

### 3.1.3 *Structural evolution versus temperature*

The evolution of the SR-XRD and NPD patterns (G4.1) upon heating are compared in Fig. 3. Below 300 K, some peaks of the SR-XRD pattern move to higher $Q$ whereas some other keep the same position versus temperature (Fig. S2 (a) and (b), Supplementary materials). This reveals an anisotropic evolution of the cell parameters. The SR-XRD and G4.1 NPD patterns were refined with a majority of monoclinic phase (LT phase) and few oxide. The small cubic oxide peaks (O) are not very sensitive to the temperature variation. Above 300 K, a structural transformation attributed to a monoclinic pseudo-cubic transition is observed (Fig. S2 (c) and (d), Supplementary material). All the patterns have been refined in the appropriate space group depending on the temperature and the measurement method (monoclinic $C2/m$ or $P2_1/c$ SG; Cubic $F$d-$3m$ or rhombohedral $R$-$3m$ SG at high T). For simplicity, we will discuss of order-disorder or monoclinic-cubic transition for the LT to HT phase transformation in the following.

The reduced cell volume (*V*/*Z*) and phase percentage are plotted in Fig. 4 upon heating and cooling. The monoclinic cell volume shows a minimum at 274 K. A mixture of monoclinic and cubic phases is observed between 325 K (315 K upon cooling) and 350 K. Both monoclinic and cubic cell volumes decrease down to 345 K and then the cubic cell volume remains stable. The O-D transition is reversible with a small thermal hysteresis. The percentage of each phase is equal to 50 % at 342(2) K. A small D desorption after heating at 360 K, can explain the smaller cell volumes upon cooling.

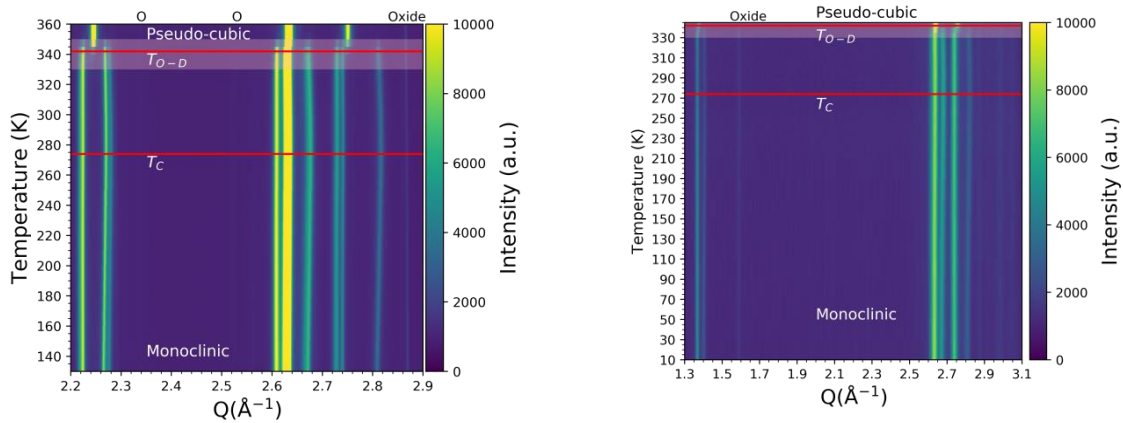

**Fig. 3**: Evolution of the diffraction patterns of $Y_{0.9}Pr_{0.1}Fe_2D_{3.5}$ measured on CRISTAL (left) and G4.1 (right) upon heating.

The cell parameter variation measured by SR-XRD and NPD is presented in Fig. 5. The $a_M$, $b_M$, $c_M$ cell parameters are described in an equivalent cubic cell ($a_M$ cubic= $\sqrt{2}/3$ $a_M$ mono, $b_M$ cubic= $\sqrt{2}.b_M$ mono, $c_M$ cubic= $\sqrt{2}.c_M$ mono) in the $P2_1/c$ description. The average cubic cell parameter is deduced from the cell volume ($a_{C1}= (2.V_M)^{1/3}$) and compared to the $a_{c2}$ belonging to the HT phase which appears at 330 K. The $a_M$ and $b_M$ parameters are extended (+0.13 % and +1.7 % at 1.5 K) whereas the $c_M$ parameter (-1.5 % at 1.5 K) is contracted compared to the average cubic parameter. Upon heating $a_M$ and $c_M$ first decrease down to 274 K and then increase again up to 330 K. Then they decrease sharply again above 330 K, i.e. in the two phase range. $b_M$ increases slightly up to 200 K then decreases again. The variation of $a_M$ and $c_M$ below 274 K is opposite to the thermal expansion behavior whereas the variation of $b_M$ is very small. These results already indicate the influence of magnetostrictive effects that will be discussed later. Above 274 K, the monoclinic cell parameter variation is rather characteristic of a reduction of the monoclinic distortion. The HT phase is refined in a pseudo-cubic cell as discussed above. The volume of the cubic cell is larger than the monoclinic one and decreases down to 345 K. This variation is also opposite to the thermal expansion and cannot be related to a deuterium desorption as it is reversible upon cooling. This also suggests a possible magnetostrictive effect.

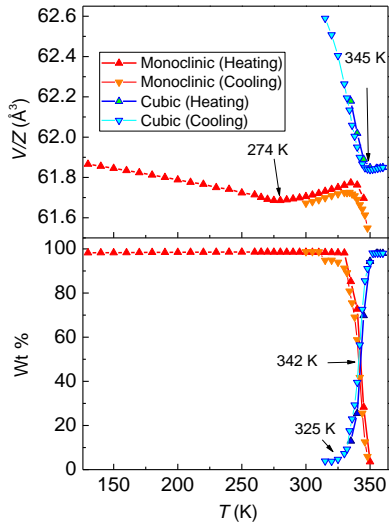

**Fig. 4**: Cell volumes and weight percentages of monoclinic and cubic phases measured by SR-XRD upon heating and cooling.

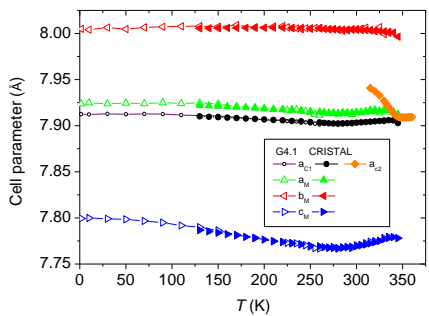

**Fig. 5**: Cell parameter variation of the monoclinic and cubic phases determined by the refinement of the NPD (open symbol) and SR-XRD (close symbols) patterns. All the parameters are described in an equivalent cubic cell as described in the text. $a_{c1}$ is an average parameter derived from the monoclinic cell volume, whereas $a_{c2}$ corresponds to the HT cubic phase.

The NPD patterns below 274 K also contains information's about the magnetic structure. The comparison of the NPD patterns shows that the two peak integrated intensities at low $Q$ (1.2 Å$^{-1}$) decreases versus temperature. Despite the noise, they show a continuous decrease down to 300 K (Fig. 6). The magnetic peaks are located at the same position than the nuclear peaks confirming that both Fe and Pr sublattices have a ferromagnetic behavior. Due to the small number of magnetic peaks, it was not possible to extract a magnetic moment for each Fe and Pr site and only an average moment can be determined from magnetic measurements.

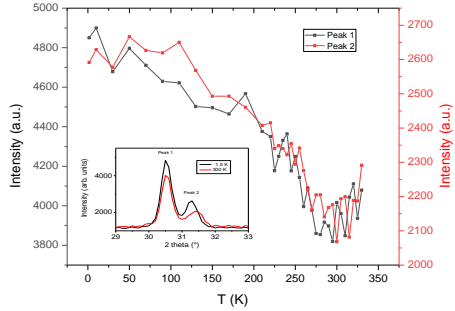

**Fig. 6**: Evolution of the magnetic peak intensities in G4.1. Inset: comparison of the two corresponding Bragg peaks at 1.5 and 300 K.

*3.3 magnetic measurements, critical behavior analysis, and magnetocaloric effect*

The magnetic measurements at 0.03 T (Fig. 7(a)) shows a ferromagnetic behavior with $T_C$= 274 K. A small magnetization increase is observed below 52 K. A decrease of the magnetization above 300 K followed by a weak peak at $T$ = 343 K is observed (Fig. 7(b)). It occurs near $T_{O-D}$ suggesting a correlation between these structural and magnetic transitions. The $M$(H) curves up to 270 K (Fig. S3 (a)) are characteristic of a ferromagnetic behavior. The magnetic cycle measured at 5 K (Fig. S3 (b)) shows a weak hysteresis loop ($\mu_0 H_C$= 200 Oe) and a saturation magnetization of 4.55(2)$\mu_B$, which results from the sum of $M_{Fe}$ (2.1 $\mu_B$/Fe) and $M_{Pr}$ (3.5 $\mu_B$/Pr). The Fe moment is larger than measured in YFe$_2$D$_{3.5}$ (1.9 $\mu_B$/Fe) [41] whereas the Pr moment corresponds to that of free ion.

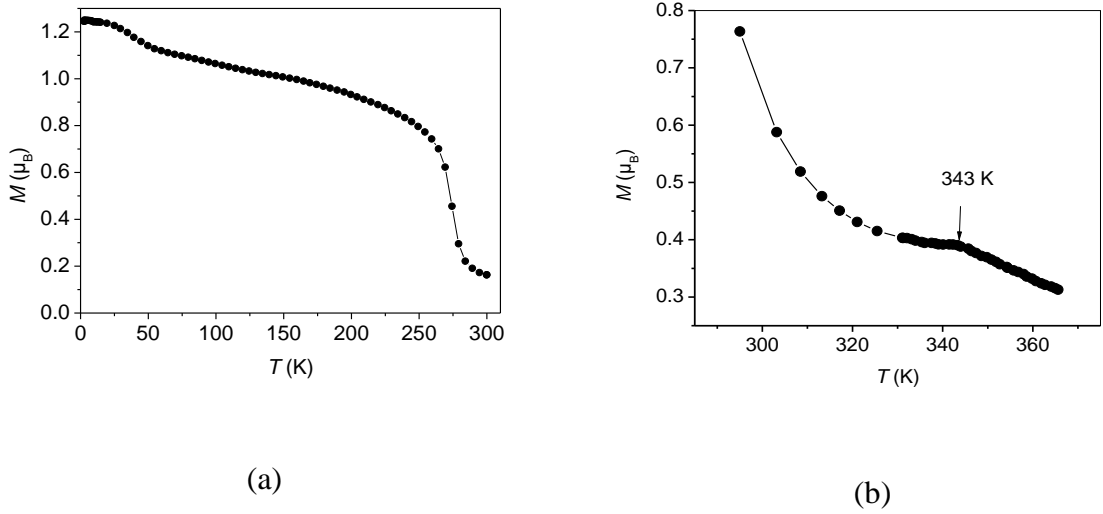

**Fig. 7**: Evolution of the magnetization curves $M$(T) at 0.05 T (a) and 0.13 T (b).

The analysis of the corresponding Arrot plots (AP) of the $M$(H) curves below 300 K shows positive slopes indicating a second order transition (Fig. S4, supplementary material). According to scaling analysis within the framework or renormalization group theory, several physical parameters for this system undergoing a second order phase transition, can be characterized by certain sets of critical exponents in the region close to the transition (details of the different models are reported in supplementary materials). The critical exponents are obtained through a well stablished and widely used iterative method [54, 55]. First the Modified Arrot Plots (MAP), where $M^{1/\beta}$ is plotted as a function of $H^{1/\gamma}$, are represented for the main universality classes, for which the classical AP is the particular case corresponding to the Mean Field model. To correctly describe the system, the MAP corresponding to that model should show good straight parallel lines at high magnetic fields in the proximity of the critical region. The straightness and parallelism of the different models has been checked for the various models provided in supplementary materials, giving both the 3D Heisenberg and 3D XY as the best starting points for the iterative procedure. The following step consists on obtaining the spontaneous magnetization and inverse of the initial susceptibility from the intersection of the straight lines in the MAP with the axes. These values are then fitted to Eqs. (1) and (2) to obtain new values of β and γ.

$$M_S(T) \sim |t|^\beta \qquad (T < T_C), \qquad (1)$$
$$\chi_0^{-1}(T) \sim |t|^\gamma \qquad (T > T_C), \qquad (2)$$

Here $t = (T - T_C)/T_C$ is the reduced temperature and β, γ are the critical exponents associated to each physical parameter. Those new values are again used to build new MAPs to check if there is an improvement on their straightness and parallelism. If this is the case, the process is repeated till the values of β and γ converge. This iterative process was performed starting from both the 3D Heisenberg and 3D XY models, and in both cases the obtained results were equivalent. The final fittings for $M_S$ and $\chi_0^{-1}$ and the final MAP are shown in Fig. 8 (a) and (b), respectively, and the critical exponents and temperatures obtained from this iterative process are collected in Table 5, where it can be seen that the critical temperature is slightly lower than the one reported previously.

Another well stablished complementary way to obtain the critical exponents is the so-called Kouvel-Fisher method [56]. This method assesses that there is a linear dependence with temperature of the spontaneous magnetization and inverse of the initial susceptibility when divided by their respective derivative with temperature, and that the slope is given by 1/β for the former case and 1/γ for the latter. The resulting fitting can be observed in Fig. 8(c), and the obtained critical exponents are listed in Table 5, with the result that the values are in very good agreement with the ones obtained from the fitting of $M_S$ and $\chi_0^{-1}$.

The last critical exponent, δ, associated to the critical isotherm, can be easily obtained as the inverse of the slope in a M vs H plot at the critical temperature ($T_C = 271$ K) in logarithmic scale according to Eq. (3):

$$M(H) \sim H^{1/\delta} \qquad (T = T_C). \tag{3}$$

The fit is shown in Fig. 8(d). Furthermore, the critical exponent δ is related to β and γ through the *Widom scaling law* ($\delta = 1 + \gamma/\beta$) [57], and, therefore, the δ obtained from the critical isotherm can be compared with the one obtained by applying *Widom scaling law* to the previously obtained β and γ values. From Table 5 it is easily seen that the values are in good agreement, confirming once again the correct evaluation of the critical exponents.

Finally, the most robust confirmation of the critical exponents is considered to be given by the correct fulfillment of the magnetic equation of state [57]:

$$M(H, t) = |t|^\beta f_\pm(H/|t|^{\beta+\gamma}), \tag{4}$$

Where $f_+$ and $f_-$ are regular analytic functions for temperatures above critical temperature and below it, respectively. A good fulfillment for this expression would imply the collapse of all the isotherms into two distinct branches (one for T>$T_C$ and the other one for T<$T_C$) for the correct β and γ values. Fig. 8(e) shows that this is in fact the case when the previously obtained critical exponents are used.

**Table 5**: Critical exponents obtained for $Y_{0.9}Pr_{0.1}Fe_2D_{3.5}$ deuteride.

| Technique | β | γ | δ |
|---|---|---|---|
| Modified Arrott Plot | 0.358 ±0.008<br>$T_C$=270.5 ± 0.2 K | 1.20 ±0.02<br>$T_C$ =270.5 ± 0.2 K | 4.3[a] ± 0.1 |
| Kouvel-Fisher Method | 0.35 ±0.01<br>$T_C$=270.4 ± 0.2 K | 1.19 ±0.01<br>$T_C$=270.5 ± 0.4 K | 4.4[a] ± 0.2 |
| Critical Isotherm | | | 4.38 ± 0.01 |

[a] Calculated from $\delta = 1 + \gamma/\beta$.

Then the magnetic entropy change $\Delta S_M$ was obtained from the magnetization isotherms using the well-known *Maxwell relation*. The analysis of the magnetic entropy variation shows negative peaks with a maxima -3.3 J kg$^{-1}$ K$^{-1}$ ($\Delta\mu_0 H$ = 4.8 T) centered at $T_C$ (Fig. S5). A weak positive peak with $\Delta S_M$ = 0.4 J kg$^{-1}$ K$^{-1}$ ($\Delta\mu_0 H$ = 1.5 T), characteristic of an inverse magnetocaloric effect is observed at 345 K (Fig. S6). The analysis of $\Delta S_M$ can be further extended by the construction of the so-called "master curves". These curves are also known as universal curves, since, for second order phase transitions, $\Delta S_M$ curves at different applied fields should collapse into a single curve in the vicinity of the critical region [58].

Master curves are obtained by normalizing $\Delta S_M$ by the value at the peak ($\Delta S_M^{pk}$) of each isotherm and rescaling the temperature axis according to [58]:

$$\theta_1 = \frac{T - T_C}{T_r - T_C} \quad (5)$$

Where $T_r$ is a reference temperature chosen as the temperature above $T_C$ for which the value of $\Delta S_M$ is half of the one at the peak. The result is shown in Fig. 9(a), and a good collapse, though not completely perfect, can be observed in the critical region, confirming the second-order nature of the transition. It has been shown that when using a single reference temperature the collapse could be distorted at T < $T_C$ due to demagnetizing field effects [59] or minor magnetic phases at higher temperatures [60].

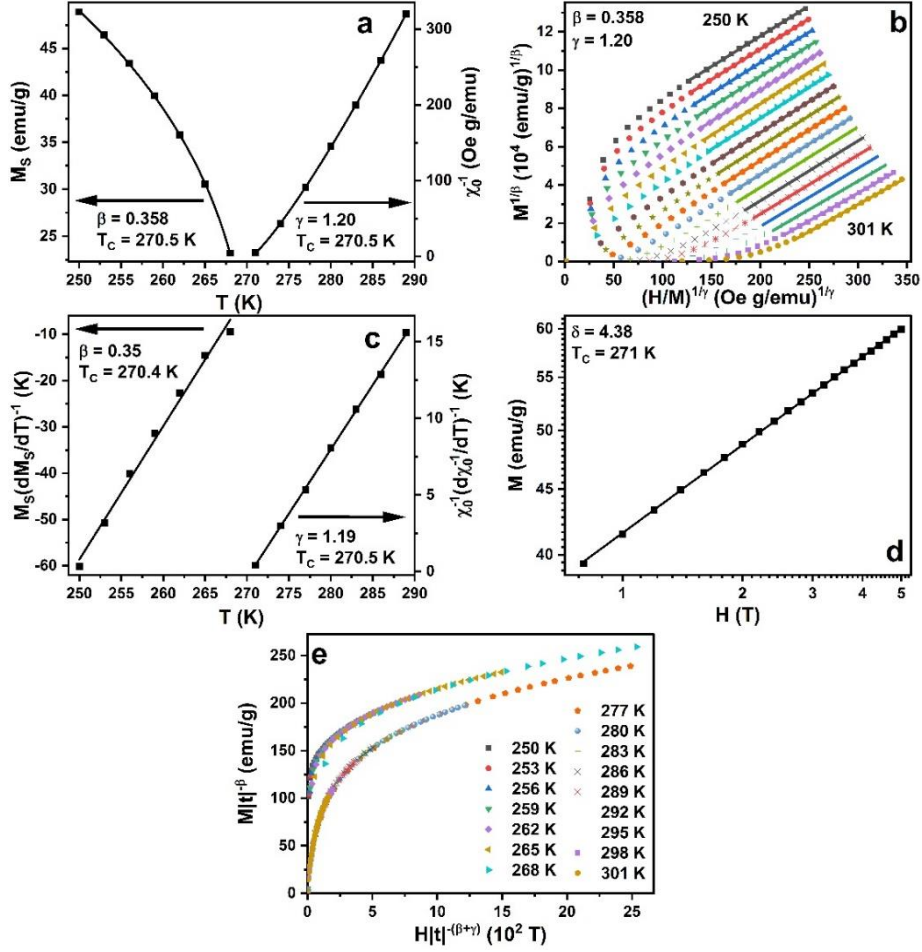

**Fig. 8:** (a) Fitting of spontaneous magnetization (left) and inverse of the initial susceptibility (right). (b) Final Modified Arrot Plot. (c) Kouvel-Fisher fitting of $M_S(dM_S/dT)^{-1}$ (left) and $\chi_0^{-1}(d\chi_0^{-1}/dT)^{-1}$ (right). (d) Critical isotherm fitting. (e) Magnetic equation of state. Lines in (a), (c) and (d) correspond to the fittings to Eqs. (1), (2) and (3), respectively.

In the former case the normalized $\Delta S_M/\Delta S_M^{pk}$ for a given temperature would increase, while in the latter case the effect would be the opposite. The inset in Fig. 9 (a) clearly shows that we are in the second case, therefore ruling out any demagnetizing field effect. The deviation is small, but could be due to the presence of the higher temperature AFM like transition which was spotted both in $M(T)$ and $\Delta S_M$ measurements. In this cases a usual approach is to rescale the temperature axis using two reference temperatures ($T_{r1}<T_C<T_{r2}$) instead of a single one [61]:

$$\theta_2 = \begin{cases} -(T-T_C)/(T_{r1}-T_C) \ , & T \leq T_C \\ (T-T_C)/(T_{r2}-T_C) \ , & T > T_C \end{cases} \quad (6)$$

In this case the collapse of the curves below the critical temperature is even better, as can be seen in Fig. 9(b).

When a material presents a second order magnetic transition several magnetocaloric effect magnitudes scale with the applied field. This is the case of the magnetic entropy change peak $\Delta S_M^{pk}$, the refrigerant capacity in its two usual definitions: as the $\Delta S_M^{pk}$ times the temperature width at half maximum ($RC_{FWHM}$), and as the area of $\Delta S_M$ enclosed between the points at half maximum ($RC_{Area}$):

$$|\Delta S_M^{pk}| \sim H^{1+(1/\delta)(1-1/\beta)} \tag{7}$$

$$RC \sim H^{1+1/\delta} \tag{8}$$

Furthermore, the reference temperature used to rescale the temperature axis in the universal curve ($T_r$) also scales with temperature as:

$$T_r \sim H^{1/(\beta+\gamma)} \tag{9}$$

where $\beta$, $\gamma$ and $\delta$ are the critical exponents we have already obtained from the critical behavior analysis.

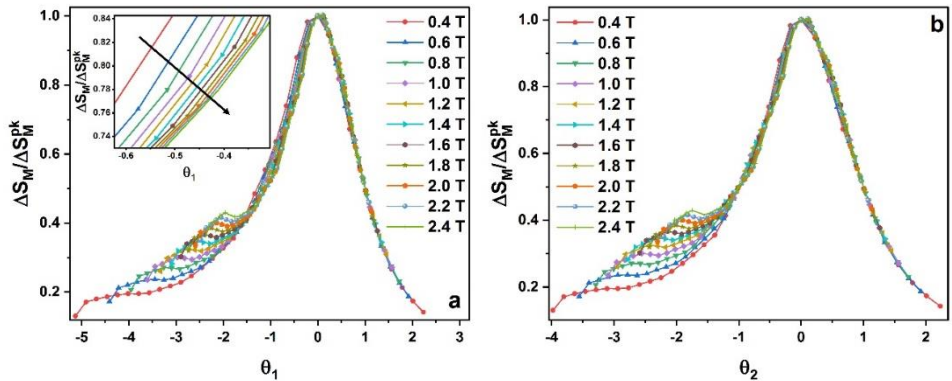

**Fig. 9**: Master curves with the rescaled temperature using one $\theta_1$ (a) and two $\theta_2$ (b) reference temperatures. The inset in (a) shows the small deviation from a perfect collapse. The direction of the arrow represents increasing applied fields.

The scaling in Eqs. (7), (8) and (9) have been successfully tested using the critical exponents in Table 5. Fig. S7 shows the results, where the red line is a visual guide to show that the experimental values lie very well along the straight line, as they should for a good scaling. This not only confirms the values obtained for the critical exponents, but also allows the extrapolation of the magnetic entropy change peak and RC values to fields that could not be experimentally accessible to compare them with literature. Furthermore Eq. (7) and (9) together with the "master curve" can be used to obtain full $\Delta S_M$ (T,H) curves at temperatures and fields which haven't been measured [62].

*3.4 DSC results*

The DSC curves for $Y_{0.9}Pr_{0.1}Fe_2D_{3.5}$ shows two reversible transitions upon heating and cooling (Fig. S8). A first one with $T_{cool}$ = 269(1) K and $T_{heat}$ = 276(1) K with $\Delta H_1$= 0.2 J/g and a second one at $T_{cool}$ = 331 K and $T_{heat}$= 338 K with an enthalpy variation of $\Delta H_2$=8.0(1) J/g. The first one corresponds to the Curie temperature ($T_C$) and the second one to the O-D transition. Both show a thermal hysteresis of 7 K. The calculated entropy variations of the transition ($\Delta S = \Delta H/T$) are $\Delta S_1$= 0.74 J kg$^{-1}$ K$^{-1}$ and $\Delta S_2$ = 24.2 J kg$^{-1}$ K$^{-1}$ corresponding to magnetic and order-disorder transitions respectively. The second one is significantly larger because it is of first order character and related to a structural transition. This is opposite to the magnetic entropy variation which was larger at $T_C$ than at $T_{O-D}$. The DSC was also measured up to 673 K, to follow the deuterium desorption. Desorption occurs above 425 K in several steps with at least three maxima at 447, 458 and 580 K. The total enthalpy of desorption is about 500 J/g, indicating a significantly larger exothermic reaction than the structural transitions. Thermal desorption, has been previously studied by in-situ neutron diffraction on $YFe_2D_x$ deuterides with various deuterium concentrations [63]. A similar multistep desorption was observed for $YFe_2D_{3.5}$ and related to the existence of deuterides with different structures and thermal stability. The temperature of desorption of each phase increases as the D concentration decreases.

**4. Discussion**

*4.1 Order-disorder transition*

The transition from monoclinic to the pseudo-cubic structure is attributed to the order-disorder transition as observed for other Laves phase hydrides and deuterides [27]. However, the high temperature phase is not fully disordered as different D sites are occupied leading to a description in a rhombohedral SG, whereas the monoclinic structure is not fully ordered as all the D sites are only partially occupied. The degree of D order in the LT phase increases with the D content as observed for $YFe_2D_{4.2}$ [29] and $ErFe_2D_5$ [64], where the occupation numbers are close to 1 or 0.5 for excluded sites.

The comparison of the cell parameters measured at 300 K by SR-XRD and refined in $C2/m$ space group shows that the Pr for Y substitution yield a slight increase of the cell parameters, as expected to its larger atomic radius: $\Delta a/a$ =0.36 %, $\Delta b/b$ =0.47 %, $\Delta c/c$ =0.34 % and $\Delta V/V$ =1.26 %. $Y_{0.9}Pr_{0.1}Fe_2D_{3.5}$ NPD pattern was refined in the same $P2_1/c$ space group, but without an additional doubling of the cell parameter as proposed for $YFe_2D_{3.5}$ [33]. As both NPD patterns show only few differences, mainly a small peak shift due to different cell parameters (Fig. S9), the previous superstructure cell used to refine the $YFe_2D_{3.5}$ pattern was probably not necessary. There is a clear structural relationship between the HT and LT phases from $R$-$3m$ to $C2/m$ and then $P2_1/c$. The lowering of crystal symmetry is mainly due to the ordering of D atoms into $R_2Fe_2$ interstitial sites. In the previous study [27] it has been observed that below 7.80 Å the filling of $R_2Fe_2$ sites is more favorable than that of $R$Fe$_3$ sites, and then the filling of $R$Fe$_3$ sites starts. Herein, the occupancy of $R$Fe$_3$ site remains less than 5 % despite the average cell volume is close to 7.9 Å. Order-disorder transitions have been observed not only for $R$Fe$_2$ hydrides, but also in several other Laves phase compounds such as $ZrCo_2$, $ZrCr_2$, $ZrV_2$, $HfV_2$, $R$Mn$_2$ with cubic or hexagonal structures. Kohlmann [27] has studied the group-subgroup scheme trees between the structure of the intermetallics and those of the corresponding hydrides or deuterides and 32 different crystal structure types, with 26 for ordered hydrides with various H content. These structures can be determined experimentally, but it remains very difficult to find a general rule to predict which Laves phase will form for a given composition.

*4.2 Ferromagnetic properties and magnetostrictive results*

The magnetic measurements show a ferromagnetic behavior with $T_C$ =274(2) K, which corresponds to the minimum observed for $a_M$, $c_M$, $\beta_M$ and $V_M$ variation versus temperature. This variation is also opposite to that expected from a thermal expansion and can only be related to a magnetostrictive effect. This change is progressive and follows also the evolution of the magnetic peak intensity. The Arrot plot variation has revealed a second order type transition. The particular variation of the monoclinic $b_M$ parameter can be related to a weak magnetostrictive effect in competition with thermal expansion. Assuming that the decrease of $a$ and $c$ upon heating is related to a decrease of the magnetic moment, is an indication that the Pr and Fe moments should be oriented in the ($a_M$, $c_M$) plane and perpendicular to the monoclinic $b_M$ axis.

Another interesting result is that $T_C$ is lower than $T_{O-D}$, meaning that the ferromagnetic order is not related to the order-disorder transition as proposed for $YFe_2D_{3.5}$ in which $T_C$ and $T_{O-D}$ were both close to 345 K [22]. Interestingly, $T_{O-D}$ is not sensitive to Pr for Y substitution, whereas $T_C$ is reduced compared to Tc=345 K forYFe$_2$D$_{3.5}$. In addition, the cell volume reduction of the high temperature pseudo-cubic phase between 315 and 345 K coincides with a magnetization decrease. This can indicate that there is also magnetostrictive effect associated to this phase, which would order at higher temperature than the monoclinic phase. In addition, the slight magnetic peak observed at 345 K, associated with a positive magnetic entropy variation indicates that there is still a correlation between both structural and magnetic transitions. As no magnetic peak is observed in the NPD pattern at 345 K, the magnetic bump should be related to a short-range magnetic order. The existence of a reverse magnetic entropy variation can be explained an elasto-caloric effect at $T_{O-D}$, as observed for austenite-martinsite transitions in Heusler compounds for example.

*4.3 Critical behavior analysis and magnetocaloric scaling*

The critical behavior analysis is a very powerful tool, which allows, by extracting the critical exponents of second order phase transitions, to explore the underlying physics after magnetic interactions and transitions, such as: the extent of the interaction, the dimension of the system or the symmetry of the spins. This analysis has already been performed in several other Laves phase compounds [65, 66], proving its interest.

The β critical exponents obtained from the fitting of the spontaneous magnetization (β=0.358 ±0.008) is between the values corresponding to 3D Heisenberg (β$_{Heis}$=0.369) and to the 3D XY (β$_{XY}$=0.348) models, while the KF analysis yields a value β=0.35 ±0.01) slightly closer to the 3D XY model. In both cases the γ values are slightly lower than the theoretical ones corresponding to any of the previous models. Deviations of the critical parameters from theoretical models can point to strong crystal field effects, magnetocrystalline anisotropy, different kind of coupling mechanisms (such as spin-orbit coupling of the itinerant electrons) or a certain degree of disorder in the system [67, 68]. The latter case is the most probable one in our system, since some disorder is still observed in the monoclinic phase due to D sites occupation. Overall, the critical exponents appear to deviate from the 3D Heisenberg isotropic universality class towards the 3D XY model, therefore presenting an easy plane of anisotropy. This is in agreement with our observations regarding the evolution of cell parameters, which suggest that the Pr and Fe moments should preferably lie in the ($a_M$, $c_M$) plane. However, as it is easily observed in Table 5, the magnetic critical exponents for these two universality classes are very close to each other, making it hard to discriminate the correct model by only using magnetic properties. In either case, the critical exponents obtained from this analysis discard long-range order interactions, for which the proper universality class would have been the Mean Field model. Therefore, short-range order magnetic interactions are responsible for the phase transition and magnetism in this system. The critical exponents have been further confirmed by checking the scaling relations of magnetocaloric properties.

## 5. Conclusions

In this work we have shown that $Y_{0.9}Pr_{0.1}Fe_2D_{3.5}$ undergoes different types of structural changes which can be correlated to magnetostrictive and magnetocaloric effects. At low temperature the deuteride crystallizes in a monoclinic structure due to a partial D order in interstitial tetrahedral sites. The compound is ferromagnetic with $T_C$ = 274(4) K. The cell parameter variation below $T_C$ is anisotropic with an increase of the $a_M$ and $c_M$ parameters upon cooling opposite to a thermal expansion behavior. This indicates that the Fe and Pr moments should be parallel to the ($a,c$) plane in agreement with the analysis of the critical exponents which point to the 3D XY model. This magnetic transition is of second order type and associated with a negative magnetic entropy variation corresponding to a direct magnetocaloric effect. Upon heating, the system undergoes a first order structural transition from monoclinic toward a pseudo-cubic phase related to a certain order-disorder transition of the D atoms in the lattice. The HT phase displays also a magnetostrictive effect with a cell volume contraction associated to a magnetization decreases. A small reverse magnetocaloric effect is observed at $T_{O-D}$ = 343 K, related to an elasto-magnetocaloric effect.

We have therefore shown that it is possible to obtain magnetocaloric effects near RT by playing on D content, although the effect remains small for an application in magnetic refrigeration. We have also demonstrated that the ferromagnetic order is not correlated to the O-D transition, contrary to what was assumed in previous study on $YFe_2D_{3.5}$, and then another type of magnetic transition occurs at $T_{O-D}$.

*Acknowledgments*

This work was done thanks to the synchrotron beam time allocated by the synchrotron SOLEIL and the neutron beam time allocated by LLB. We are also grateful to Florence Porcher and Francoise Damay for their help as local contact for neutron diffraction measurements. A. Herrero thanks the Department of Education of the Basque Government as grantee of the

**Supplementary materials**

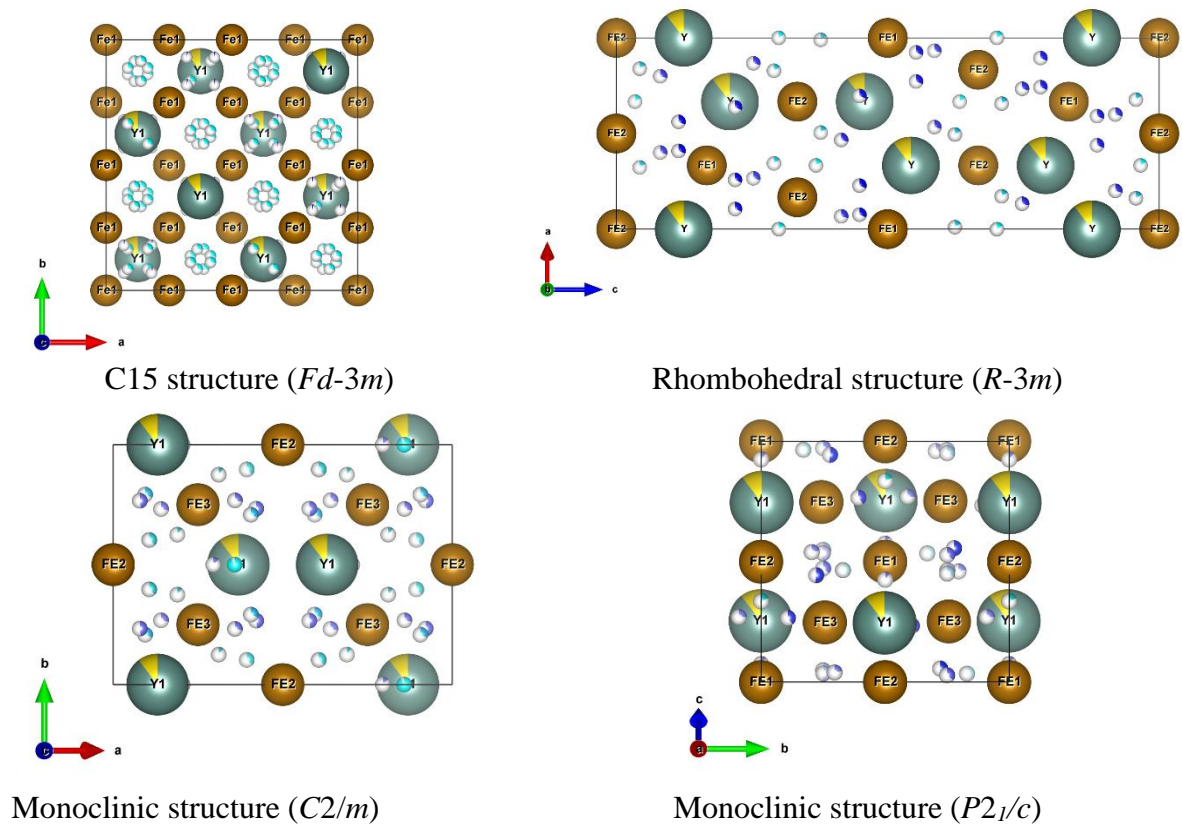

C15 structure (*Fd-3m*)

Rhombohedral structure (*R-3m*)

Monoclinic structure (*C*2/*m*)

Monoclinic structure (*P*2$_1$/*c*)

**Fig. S1**: Structure representation of Y$_{0.9}$Pr$_{0.1}$Fe$_2$D$_{3.5}$ in the different space groups

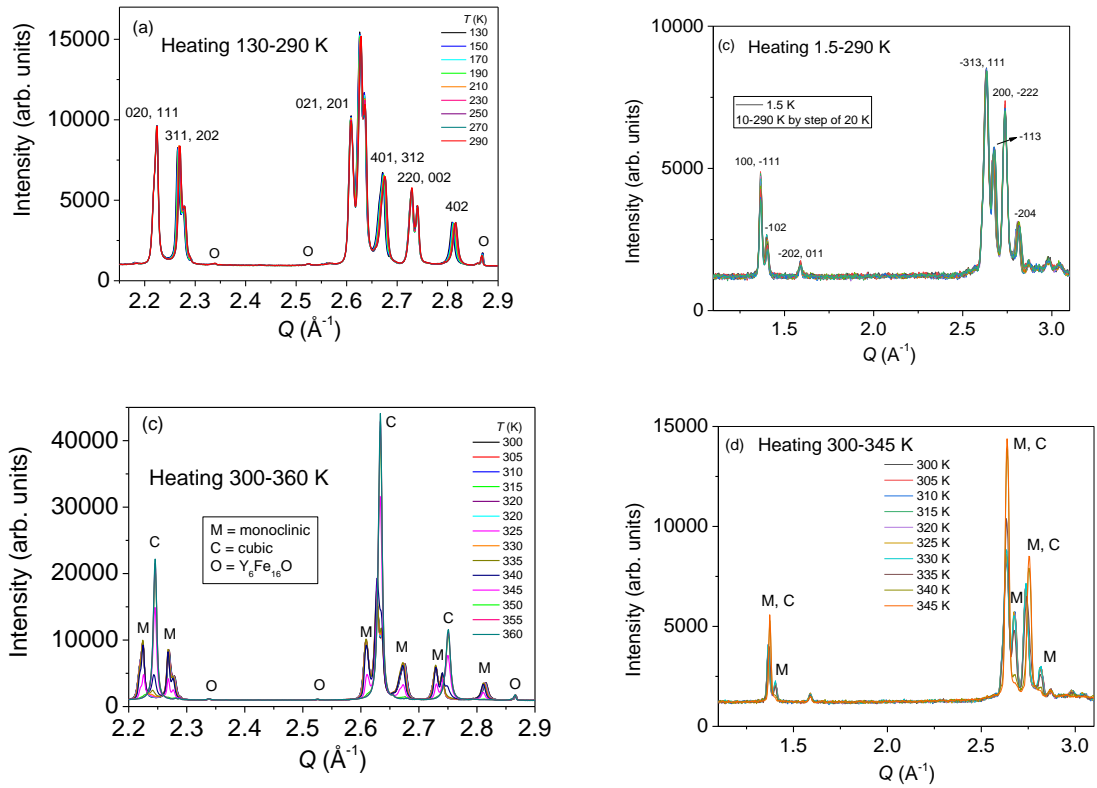

**Fig. S2**: Comparison of the diffraction patterns of $Y_{0.9}Pr_{0.1}Fe_2D_{3.5}$ measured on CRISTAL (a, c) and G4.1 (b, d) at different temperatures below (a, b) and above (c, d) 300 K. All the presented patterns were measured upon heating.

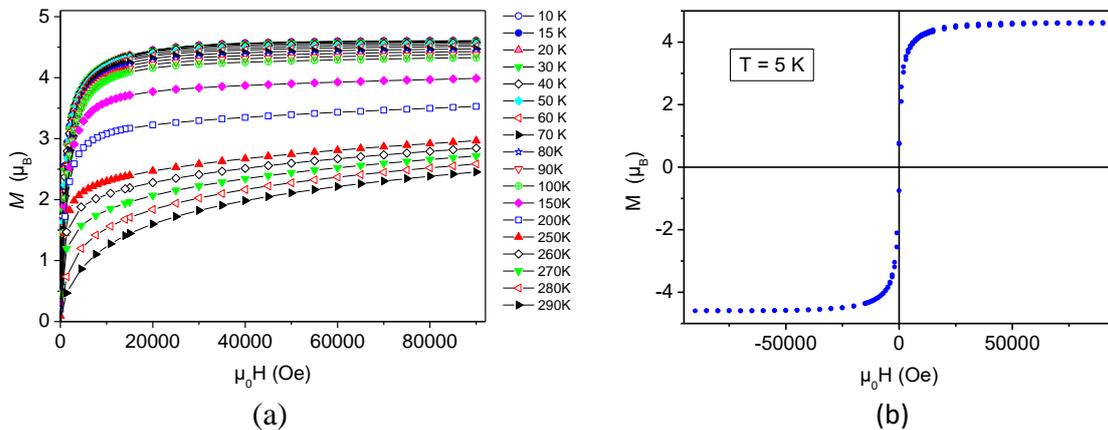

**Fig. S3**: Magnetization curves versus applied field at different temperatures and hysteresis cycle at 5 K.

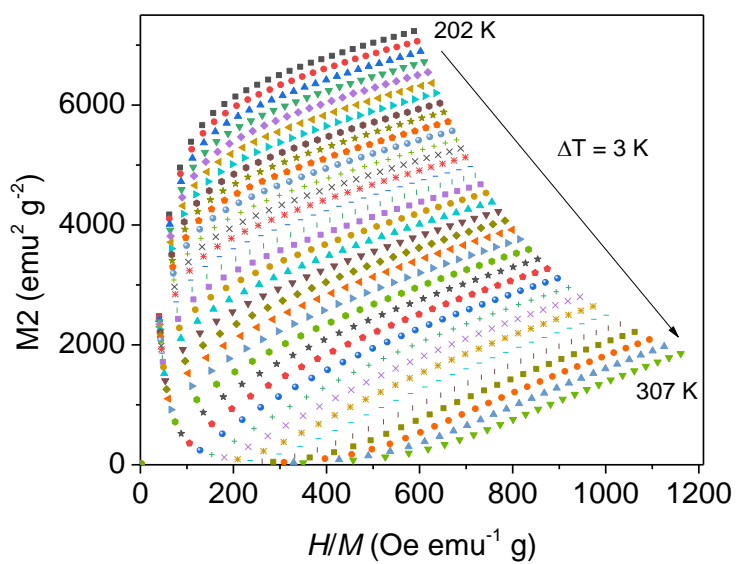

**Fig. S4**: Arrot Plots for $Y_{0.9}Pr_{0.1}Fe_2D_{3.5}$ deuteride

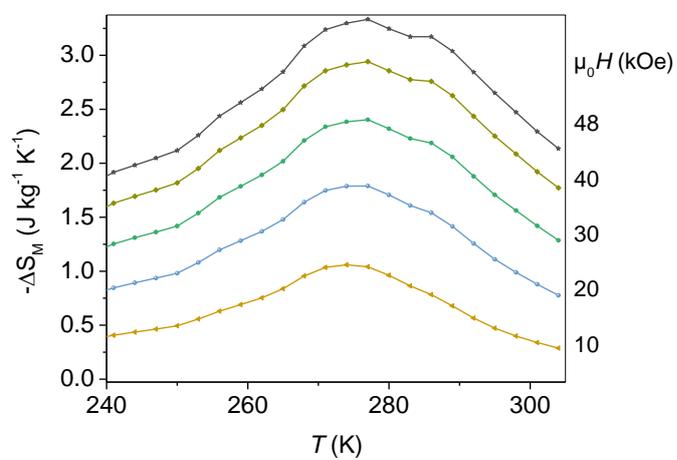

**Fig. S5** : Evolution of the magnetic entropy variation curves around $T_C$ at different applied field.

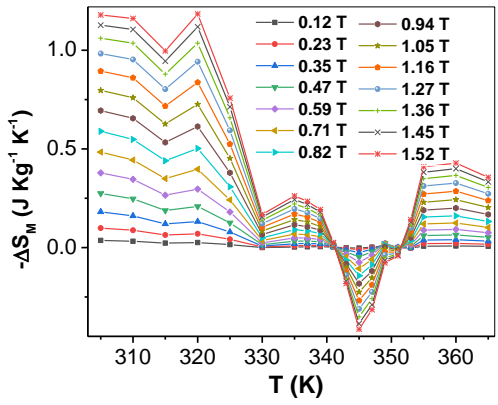

**Fig. S6**: Evolution of the magnetic entropy variation curves above room temperature.

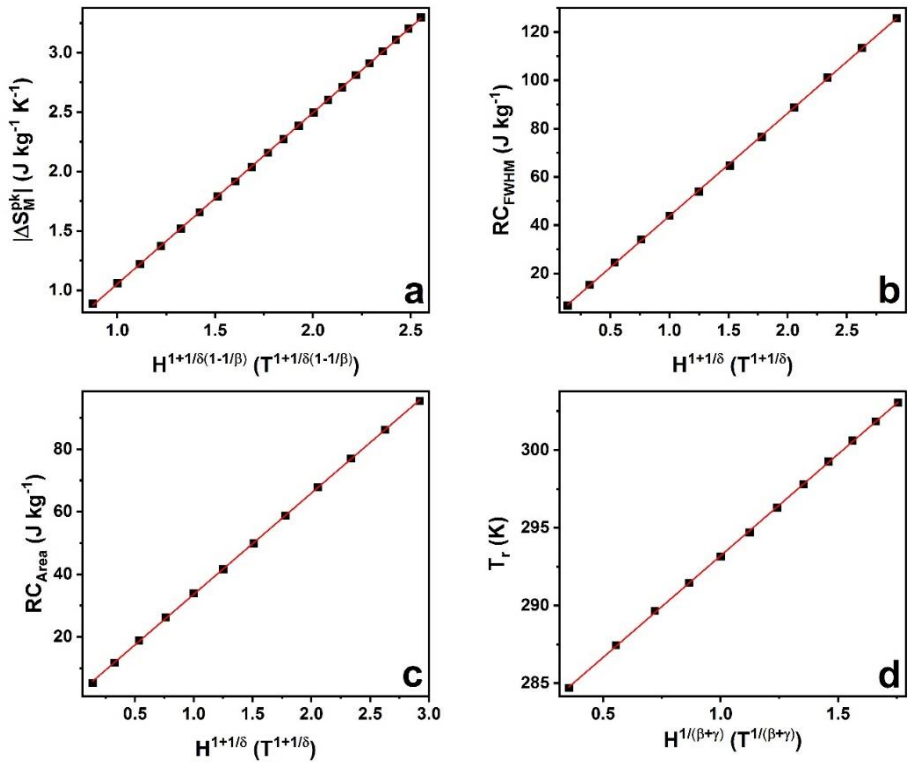

**Fig. S7:** Scaling relations for: the magnetic entropy change peak (a), $RC_{FWHM}$ (b), $RC_{Area}$ (c), and reference temperature $T_r$ (d). $\beta$, $\gamma$ and $\delta$ values are the ones gathered in Table 6. The red line is a visual guide to check that the experimental data lies on a straight line.

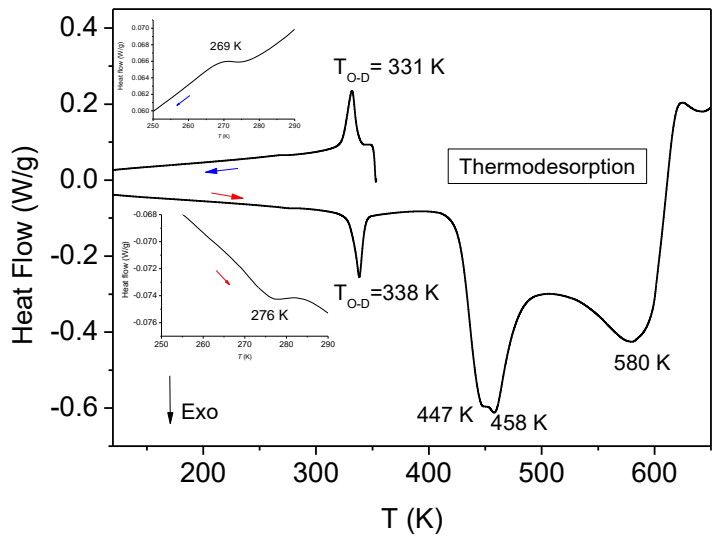

**Fig. S8** : DSC curves versus temperature for $Y_{0.9}Pr_{0.1}Fe_2D_{3.5}$ showing different structural transitions and thermal desorption. In inset: zoom around 270 K, to see more clearly the DSC peak.

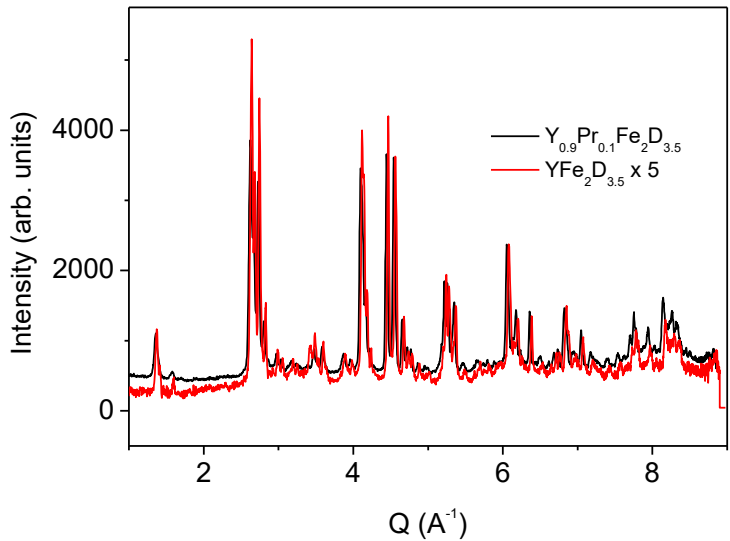

**Fig. S9**: Comparison of the NPD patterns of $YFe_2D_{3.5}$ (unpublished data from (Wiesinger, 2005)) and $Y_{0.9}Pr_{0.1}Fe_2D_{3.5}$

**Determination of critical exponents**

According to scaling analysis within the framework or renormalization group theory, several physical parameters of systems undergoing a second order phase transition can be characterized by certain sets of critical exponents in the region close to the transition. This is the case, for example, of the spontaneous magnetization ($M_S$), inverse of the initial susceptibility ($\chi_0^{-1}$) and critical isotherm [1]:

$$M_S(T) \sim |t|^\beta \qquad (T < T_C), \qquad (S1)$$

$$\chi_0^{-1}(T) \sim |t|^\gamma \qquad (T > T_C), \qquad (S2)$$

$$M(H) \sim H^{1/\delta} \qquad (T = T_C). \qquad (S3)$$

Here $t = (T - T_C)/T_C$ is the reduced temperature and β, γ, δ are the critical exponents associated to each physical parameter. These critical exponents are predicted for several universality classes, which are obtained from the development of different Hamiltonians, and therefore, the attribution of a given material to a certain universality class allows the understanding of the underlying physical mechanisms responsible for the magnetic interactions in the system. The most common models for magnetic materials are gathered in Table S1. They can be divided into two groups according to their magnetic interaction range. If long-range order interactions are responsible for the magnetic ordering, the Mean Field model would be of application, while the rest describe different short-range order spin alignments: 3D Heisenberg describes an isotropic system, 3D XY describes a system where there is an easy plane of magnetization, and 3D Ising model describes a system with uniaxial anisotropy.

**Table S1**: Set of critical exponents for the most common universality classes in magnetic materials.

| Universality class | $\beta$ | $\gamma$ | $\delta$ | Ref. |
|---|---|---|---|---|
| Mean-field Model | 0.5 | 1.0 | 3.0 | [2] |
| 3D-Ising | 0.3265 | 1.237 | 4.79 | [3] |
| 3D-XY | 0.348 | 1.317 | 4.78 | [4] |
| 3D-Heisenberg | 0.369 | 1.396 | 4.78 | [5] |